\documentclass{article}
\usepackage{spconf,amsmath,graphicx}
\usepackage{multirow}
\usepackage{threeparttable}
\usepackage{color,xcolor}
\usepackage{booktabs}
\usepackage{float}
\usepackage[section]{placeins}
\usepackage{cite}

\title{PL-EESR: PERCEPTUAL LOSS BASED END-TO-END ROBUST SPEAKER REPRESENTATION EXTRACTION}
\name{Yi Ma\textsuperscript{1}, Kong Aik Lee\textsuperscript{2}, Ville Hautam\"aki\textsuperscript{3}, Haizhou Li\textsuperscript{1}\thanks{This research / project is supported by Science and Engineering Research Council, Agency of Science, Technology and Research (A*STAR), Singapore, through the National Robotics Program under Human-Robot Interaction Phase 1(Grant No. 192 25 00054), its RIE 2020 Advanced Manufacturing and Engineering Human (AME) Programmatic Grant (Grant No. A18A2b0046) and its Feasibility Study Scheme (Project No. FS-2021-001).}}
\address{\textsuperscript{1}Department of Electrical and Computer Engineering, National University of Singapore, Singapore\\
\textsuperscript{2}Institute for Infocomm Research, A*STAR, Singapore\\
\textsuperscript{3}School of Computing, University of Eastern Finland, Finland}
\begin{document}
%
\maketitle
\begin{abstract}
Speech enhancement aims to improve the perceptual quality of the speech signal by suppression of the background noise. 
However, excessive suppression may lead to speech distortion and  speaker information loss, which degrades the performance of speaker embedding extraction. To alleviate this problem, we propose an end-to-end deep learning framework, dubbed PL-EESR, for robust speaker representation extraction. This framework is optimized based on the feedback of the speaker identification task and the high-level perceptual deviation between the raw speech signal and its noisy version. We conducted speaker verification tasks in both noisy and clean environment respectively to evaluate our system. Compared to the baseline, our method shows better performance in both clean and noisy environments, which means our method can not only enhance the speaker relative information but also avoid adding distortions. 
\end{abstract}
\section{Introduction}
\label{sec:intro}
%
%
Speaker embedding refers to a fixed-length continuous-valued vector extracted from a variable-length utterance~\cite{lee2021asvtorch}. In speaker verification, the extracted speaker embedding is forwarded to the backend classifier. It is important that a speaker embedding charaterizes well the speaker's individuality. 
Ideal speaker embedding contains solely the speaker information so that extracted speaker embeddings are similar for the same speaker, and very different between speakers. Traditionally,  GMM supervectors~\cite{campbell2006support,kenny2008study} and i-vector~\cite{dehak2009support} were used to extract speaker embedding. With the development of deep learning, d-vector~\cite{variani2014deep} and x-vector~\cite{snyder2018x} were proposed recently. The basic hypothesis of all of these extraction methods is that the speaker embedding contains solely one speaker’s information. However, in the presence of background noise and interference, it is difficult to record only the voice of the interested speaker.
\begin{figure}[!t]
\centering
\includegraphics[trim=0cm 0cm 0cm 0cm, clip, width=9cm]{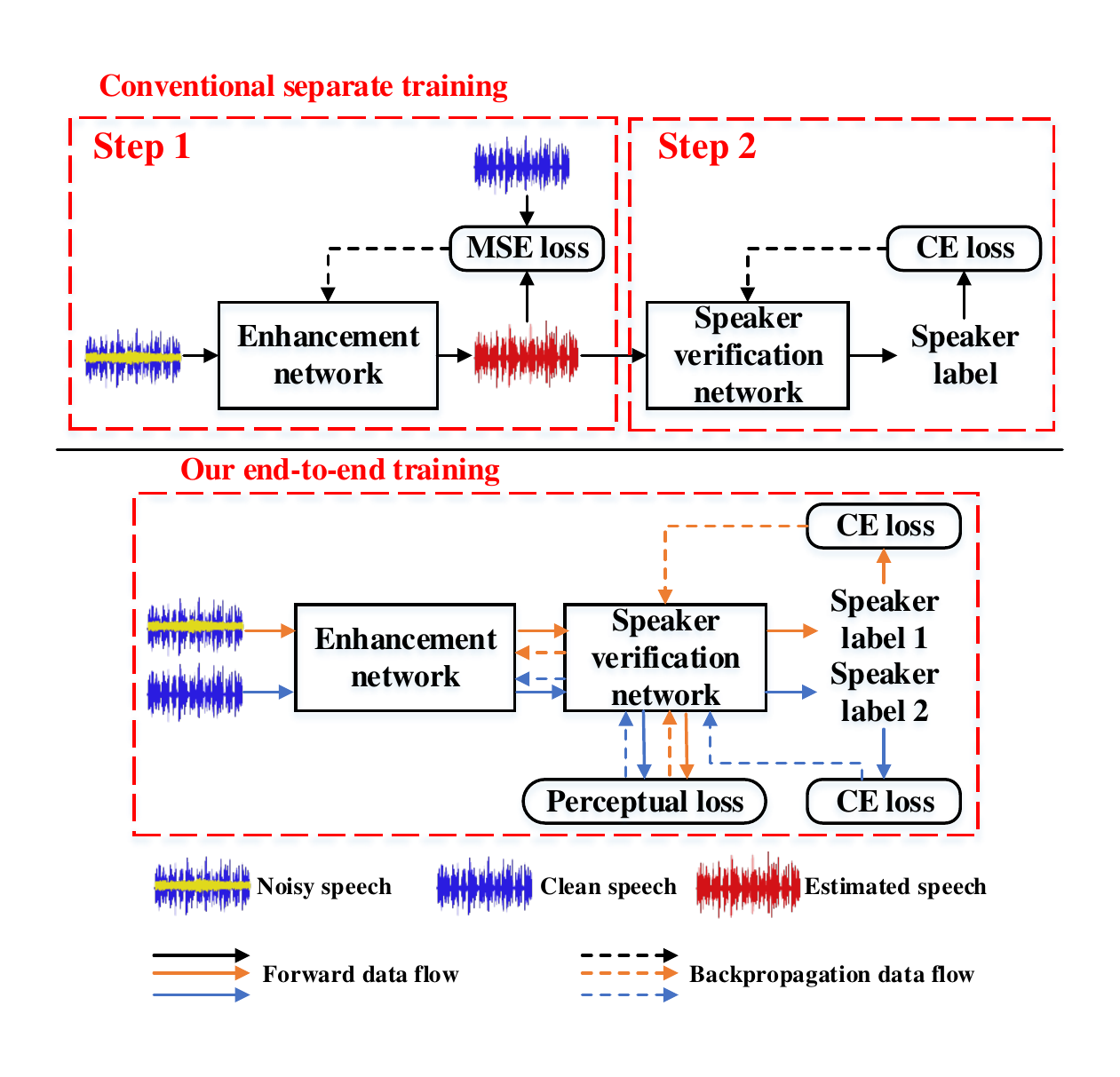}
\caption{A diagram illustrating noise-robust speaker representation using a separate training scheme (top panel), and an end-to-end training scheme (bottom panel).}
\label{fig:Overview}
\end{figure}
In this work, we seek to extract consistent speaker embedding for a monaural utterance under either noisy or clean conditions.

Speech enhancement, which aims to keep the target signal of interest and filter out additive background noise~\cite{loizou2007speech}, is a conventional method for handling noisy utterances.
Recently, deep learning based methods have shown significant improvement over the classical methods. 
These methods usually generate a mask imposed element-wise on the original signal to estimate the underlying clean signal \cite{kjems2009role,wang2014training,yu2020constrained,williamson2015complex,pandey2020learning,liu2020speaker}.
%
Usually the model is trained using mean square loss on noisy samples and targets. This function only guarantees proximity of the enhanced speech to clean speech and simply accounts for the averaged errors, which leads to artifacts in the output.
It has been observed that speaker recognition accuracy is often hurt by first enhancing the speech signal and then performing the recognition~\cite{sadjadi2010assessment,shi2020robust}. 
%
%
%

In addition to using classification loss, we can use also task-specific loss, such as  {\em perceptual loss}, or also called {\em deep feature loss}~\cite{johnson2016perceptual}. Perceptual loss is based on the difference of high-level feature representations extracted from a pre-trained auxiliary network. 
It was first proposed to deal with the image style transfer and super-resolution task~\cite{johnson2016perceptual}. 
%
This idea can be generalized for example to 
%
speaker verification task-specified enhancement training~\cite{kataria2020feature,kataria2020analysis}. The approach used in~\cite{kataria2020feature,kataria2020analysis} is similar to our method to compare the activation of enhanced and referenced signals using the auxiliary network. 
However, one drawback of \cite{kataria2020feature,kataria2020analysis} is that the auxiliary network is different from the speaker representation network although these two models are trained using the same task. The number of the total parameters in their system is up to 26.1M, which is larger than a standard x-vector network. 
This also means that they adopt a two-stage training strategy and the enhancement model is independent of the speaker verification model. 
%
Another drawback is that the perceptual loss used in \cite{kataria2020feature,kataria2020analysis} relies heavily on clean speech as a training target which we need to select manually.
In this paper, we propose an end-to-end joint training network, of which the total number of parameters is 9.8M. And we modified the perceptual loss to make it work for clean utterances as well.

Recent research suggests an end-to-end scheme that combines the speech enhancement task with another downstream speech task \cite{hou2020multi} because it reduces the distortion.
As for joint training of speech enhancement with speaker verification, similar work is \cite{shon2019voiceid}, in which the enhancement module is trained using the loss function of speaker identification task to improve the accuracy of speaker verification in both clean and noisy conditions.
Although their method shows improvement for speaker verification, intuitively the proposed VoiceID loss based on softmax only tries to enlarge the inter-class differences among different speakers.
However, a robust speaker representation network should not only maximize the inter-class distance but also minimize the intra-class variations of the learned embeddings. 
For this reason, we proposed the PL-EESR model, which focused on both inter-class and intra-class distance efficiently.

The contribution of this paper is twofold. Firstly, we proposed a robust end-to-end speaker representation network optimized using perceptual loss and cross entropy loss.
%
%
Our model contains two parts: a task-specific enhancement module and a speaker embedding extraction module.
During training, firstly these two modules are pre-trained in proper order: the embedding extraction module is trained on speaker identification task and fixed first; then the enhancement module is trained using cross entropy loss and perceptual loss with the embedding extraction module as an auxiliary network.
%
Secondly, these two modules are fine-tuned simultaneously to reduce potential mismatch. The idea is illustrated in Fig. \ref{fig:Overview}
%
%
To verify the effectiveness of our proposed network, we performed a speaker verification task on the development and evaluation part of Speaker In The Wild(SITW). The SITW is viewed as a high SNR condition in our work. We also simulated a noisy SITW by corrupting SITW with background noise. This noisy SITW is set as low SNR condition.

\section{perceptual loss}
\subsection{Problem description}
Let us denote the $X$ to be the log-mel spectrogram of a noisy utterance. 
The utterance is corrupted by the additive background noise so that it can be written as $X=S+N$, where $S$ and $N$ are log-mel spectrogram of the clean utterance and the noise.
The target of speech enhancement in our work is to estimate a mask $M$ and we can get the estimated clean spectrogram $\hat{S}$ by the element-wise multiplication of the mask and the input utterance: $\hat{S}=X\otimes M$.

Conventionally, a speech enhancement network is trained using the Euclidean distance between the estimated clean spectrogram and the ground-truth: $\mathcal{L}(S,\hat{S})= \rVert S-\hat{S}\rVert_2$.
However, this method may cause some speaker related information loss or distortion in the resultant enhanced spectrogram. 
If we apply the enhanced spectrogram on the speaker embedding extraction task directly, it may cause degraded performance, especially for high SNR conditions. 
We believe this is because the Euclidean distance can not capture the perceptual difference between estimated and ground-truth spectrogram.

\newcommand{\tabincell}[2]{\begin{tabular}{@{}#1@{}}#2\end{tabular}}
  \begin{table}[!t]

  \caption{Dataset and loss function for each stage.\\}
  \label{dataset}
  \centering
  
  \begin{tabular}{ @{}cccc@{} }

    \toprule
    \textbf{Stage}& \textbf{Clean data}& \textbf{Noisy data}& 
    \textbf{Loss} \\
    \midrule
    Pre-training 1 & \textbf{vox\_clean\_aug} & - & CE loss\\
    \midrule

    \multirow{5}*{\tabincell{c}{Pre-training 2;\\ Finetune}} & \textbf{vox\_clean} & \textbf{vox\_noisy} & \multirow{5}*{\tabincell{c}{CE loss;\\ Percep\\-tual loss}}  \\
     & \textbf{vox\_clean} & \tabincell{c}{\textbf{vox\_noisy\_aug}\\\textbf{(MUSAN)}} &  \\
     & \tabincell{c}{\textbf{vox\_clean\_aug}\\\textbf{(RIRs)}} & \tabincell{c}{\textbf{vox\_noisy\_aug}\\\textbf{(RIRs)}} &  \\
  
    \bottomrule
  \end{tabular}

\end{table}
\subsection{Perceptual loss}
It is believed that a trained deep network shows different activations in their hidden layers, which helps the model to learn characteristics from the input feature. Based on this assumption, the perceptual loss, which is the distance of activations in hidden layers of the trained auxiliary network with estimated and ground-truth spectrogram as input respectively, is proposed in~\cite{johnson2016perceptual}.

According to the principle of perceptual loss, the auxiliary network decides what information to be kept. In this paper, our goal is to enhance the speaker relative information.
Therefore, the speaker verification network can be trained first, then it is fixed as the auxiliary network to extract perceptual loss for the enhancement module:
\begin{equation}
    \mathcal{L}(X)= \sum_{i=1}^j\rVert \phi_i(X)-\phi_i(\theta(X))\rVert_2
    \label{perceptual}
\end{equation}
where the $\theta$ and the $\phi$ are the  enhancement module and auxiliary speaker embedding extraction module respectively, and $\phi_i$ denotes the $i_{th}$ layer of embedding extraction module. Equation (\ref{perceptual}) is what \cite{kataria2020feature, kataria2020analysis} used.

\section{METHOD}
\subsection{Architecture}
\label{architecture}
We now introduce the end-to-end model that incorporates enhancement process and speaker representation together. The flow chart is illustrated in Fig. \ref{exp} (d).

As shown in Fig. \ref{exp} (d), at first, the feature (30-dimensional log-Mel spectrogram) is extracted from the speech signal in the time domain. It was used for both the enhancement module and the representation module. 

To remove the effect of the corpus channel, as mentioned in \cite{pandey2020cross}, we applied the channel normalization to help the enhancement module converge rapidly. The mean $\mu$ and standard deviation $\sigma$ of the clean and noisy channel are computed using the clean and noisy training set respectively. 
Noisy utterance is normalized before enhancement, as follow:
\begin{equation}
    X(k) \leftarrow \frac{X(k)-\mu_n(k)}{\sigma_n(k)^2}
\label{normalization}
\end{equation}
An inverse normalization is then performed on the enhanced speech:
\begin{equation}
    \hat{S}(k) \leftarrow \hat{S}(k)\times \sigma_c(k)^2+\mu_c(k)
\label{inverse-normalization}
\end{equation}
where $\sigma_n$ and $\mu_n$ are the statistical parameters of the noisy training set, and $\sigma_c$ and $\mu_c$ are corresponding clean version. The $k$ denotes the index of Mel-filter and $X(k)$ is the feature corresponding to $k_{th}$ Mel-filter. After the inverse-channel-normalization, instance normalization is applied before speaker representation extraction:
\begin{equation}
    \hat{s}_i(k) \leftarrow \frac{\hat{s}_i(k)-\mu_i(k)}{\sigma_i(k)^2}
\label{ins-normalization}
\end{equation}
where $\sigma_i$ and $\mu_i$ are the statistical parameters of each utterance $\hat{s}_i$. 

To exploit the temporal context information effectively, three bidirectional long short-term memory (BLSTM) layers and one fully-connected layer are stacked as the enhancement module. The number of features in the hidden state of BLSTM is 128. The activation of the sigmoid is applied on the output of fully-connected layer to generate the mask in the range 0 to 1.

Our verification module adopts the x-vector. The first five TDNN layers  extract frame-level information. Then two fully connected layers extract information in utterance-level. The two phases are connected by a global average pooling layer. Since we don't know the effect of activation in each layer on our task, all of the five TDNN layers and the first fully connected layer are used to compute perceptual loss in the training stage. The speaker embedding was extracted from the first fully connected layer during inference. 

\subsection{Training strategy}
\begin{figure*}[!t]
\centering
\includegraphics[trim=0cm 0.5cm 0cm 0cm, clip, width=18.3cm]{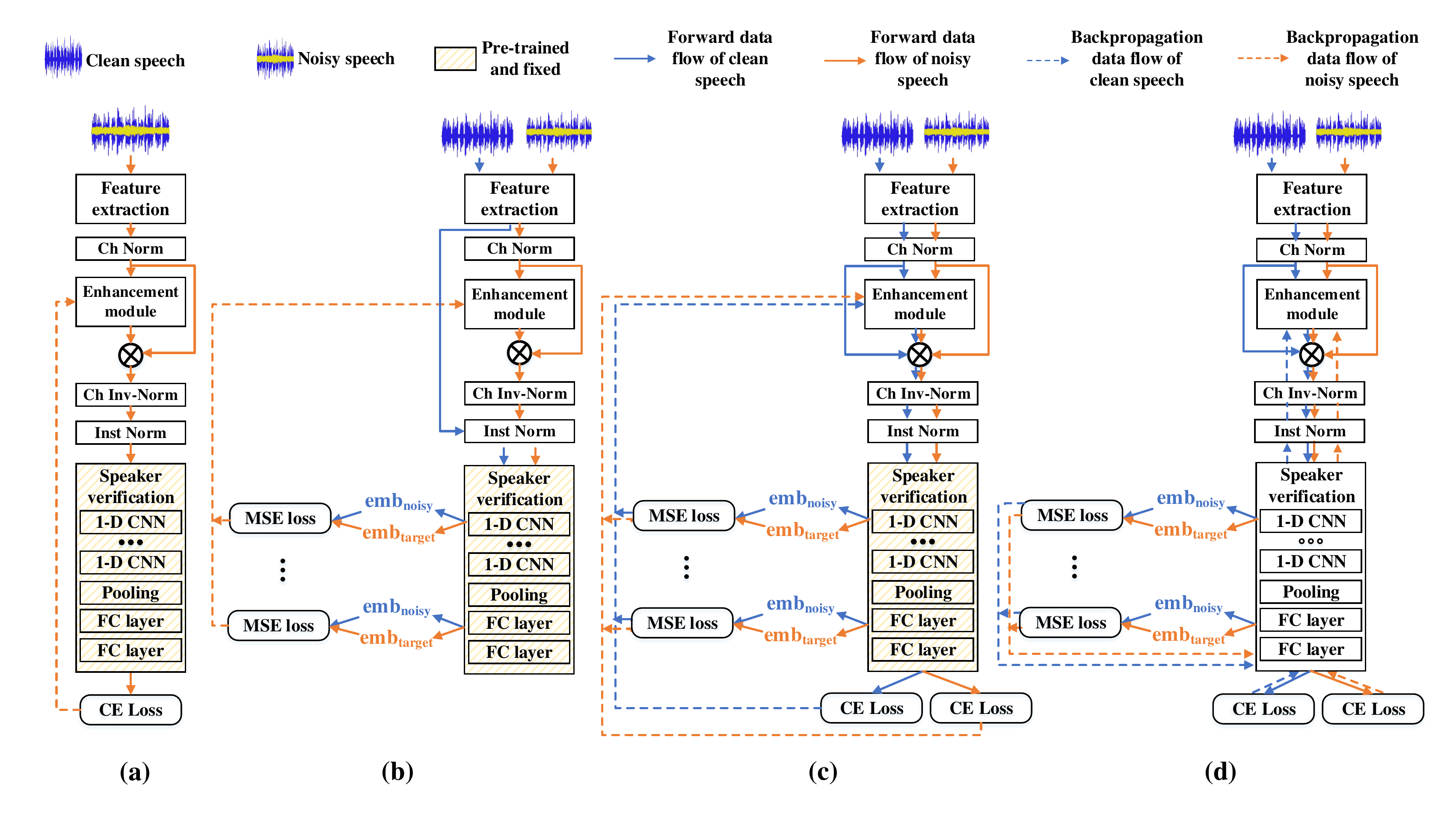}
\caption{Block diagrams of (a) optimizing speech enhancement module using cross entropy loss; (b) optimizing speech enhancement module using perceptual loss on noisy speech; (c)optimizing speech enhancement module using cross entropy loss and perceptual loss on both noisy and clean speech; (d)optimizing speech enhancement and speaker extraction modules jointly using cross entropy loss and perceptual loss on both noisy and clean speech. The ``Ch Norm" and ``Ch Inv-Norm'' denote channel normalization and channel inverse-normalization, ``Inst Norm" is the instance normalization.}
\label{exp}
\end{figure*}

The perceptual as in (\ref{perceptual}) was used in \cite{kataria2020feature} to train the speech enhancement network. 
The idea is illustrated in Fig. \ref{exp}(b): the clean utterance is used as the target and the noisy utterance is trained to show the same activations of the auxiliary network as the target does.
However, the gradient of (\ref{perceptual}) only concerns the enhancement for noisy utterance. It may cause some potential detrimental to the clean utterance.
Therefore, we propose the training scheme as in Fig. \ref{exp}(c) and (d): the mask is generated and applied on clean utterance as well, and the enhancement module is optimized using the gradient of both noisy utterance and clean utterance.
The goal of the enhancement module is to make sure the noisy utterance and the clean utterance have a consistent representation. 
So the perceptual loss in our system is modified as:
\begin{equation}
        \mathcal{L}_{pcptl}(\theta(X),\theta(S))= \sum_{i=1}^j\rVert \phi_i(\theta(X))-\phi_i(\theta(S))\rVert_2
\end{equation}
where the $X$ and $S$ are the noisy and clean utterance respectively. As mentioned in Section \ref{architecture}, $j$ equals $6$ in our case. 

A good speaker representation network should minimize the distance between the utterances belonging to the same speaker, while maximize the distance of different speakers in embedding vector space.
The cross entropy loss in (\ref{ce}) is a common loss function in the classification problem to enlarge the distance of different speakers. 
\begin{equation}
    \mathcal{L}_{ce}(\phi(\theta(S)),y)=-log\frac{exp(\phi(\theta(S))[y])}{\sum_{y'}exp(\phi(\theta(S))[y'])}
\label{ce}
\end{equation}
In our case, the $\phi(\theta(S))$ is the speaker label estimated by the speaker representation module following the speech enhancement process, and $y$ is the ground-truth speaker label.
Intuitively, the softmax in (\ref{ce}) enlarge the intra-class discrimination, but it shows no effect on inter-class distance. 
To remedy this limitation, we train our system using cross entropy loss and perceptual loss jointly:
\begin{equation}
    \mathcal{L}=\lambda\times\mathcal{L}_{pcptl}+(1-\lambda)\times\mathcal{L}_{ce}
\end{equation}
where the perceptual loss can be viewed as the criterion for inter-class distance. The $\lambda$ is a constant for balancing the intra-class and inter-class distance. We set the value of $\lambda$ as 0.5 in our work. (Code is available online\footnote{Source code: https://github.com/mmmmayi/PL-EESR}.)

%
%
%
%

\section{Experiments}
\subsection{Datasets}
\label{data}
\textbf{Training set:} we combine VoxCeleb1 and VoxCeleb2 \cite{chung2018voxceleb2} as our training set. Background noise in this dataset is inevitable because it is collected from YouTube. But we still set samples in this dataset as clean utterances in our experiment and we name it as \textbf{vox\_clean}. It should be noted that speakers who are in both VoxCele2 and Speakers in the Wild (SITW) \cite{mclaren2016speakers} are removed from voxceleb2 because we use the SITW as our test set. 
The \textbf{vox\_clean} is augmented with noise drawn from MUSAN \cite{snyder2015musan} and is convolved with simulated RIRs \cite{ko2017study}. The augmentation process is based on Kaldi SITW/v2 recipe \cite{povey2011kaldi}. The augmented \textbf{vox\_clean} is called \textbf{vox\_clean\_aug}. 
%
To simulate speech in noisy conditions, we corrupt the \textbf{vox\_clean} using the noise randomly selected from MUSAN as noisy utterance. The SNR of each noisy utterance is randomly selected from the range of 0 to 20 excluding $\{0,5,10,15,20\}$. This data set is named \textbf{vox\_noisy}. Each sample in the \textbf{vox\_noisy} has a corresponding clean version in the \textbf{vox\_clean}.
To be consistent with the \textbf{vox\_clean\_aug}, the \textbf{vox\_noisy} is augmented in the same procedure and we call it \textbf{vox\_noisy\_aug}.\\
\textbf{Test set:}
We use two datasets to evaluate the performance of our method in noisy and clean conditions. The first dataset is SITW. Being similar to VoxCeleb, the speech in SITW is collected from open-source media channels as well.  We use the development and evaluation part core-core condition of SITW to evaluate our network in clean condition. 
We generated the noisy Speakers in the Wild (NSITW) as our second test set. The noise used to corrupt the SITW is provided by DNS-challenge \cite{reddy2021icassp}. We manually select the noise categories which are similar to MUSAN (bubble, music and plain noise). Totally 282 categories are used and the SNR of each noisy utterance in NSITW is randomly selected from set $\{0,5,10,15,20\}$. The NSITW is used to evaluate the performance of our model in noisy conditions.

\subsection{Feature}
In VoxCeleb1 and VoxCeleb2, multiple scenarios were given for each speaker, and there are several utterances in each scenario. We concatenated utterances in the same scenario to one training utterance. We generate 30-D log Mel-spectrogram as the feature from each training utterance. The sample rate of all utterances in our work is 16kHz. Our feature is extracted with a Hann window of 400 frames width and 160 frames hop size. This feature is used for both speech enhancement module and speaker representation module, but we need to note that mean and variance normalization (MVN) is performed only for speaker representation. The training utterance whose length is less than 500 frames and the speaker whose utterance amount is less than 10 are removed. Therefore, there are 5916 speakers in the training set. We randomly clipped a consecutive 300-frame segment from each training utterance to optimize our network during training. 
We do not use voice activity detection in the training stage because the silence segment has been removed from VoxCeleb1 and VoxCeleb2, but energy based voice activity detection is performed in the test stage. 

\subsection{Training}

\label{training}

Our training process includes two stages: pre-training stage and finetune stage.

For the pre-training stage, firstly, we trained our speaker representation module (Pre-training 1) with a batch size of 512 and optimizer of stochastic gradient descent (SGD). The initial learning rate is 0.2 and it decreases by half when the loss decrease ratio in each epoch is less than 0.01. The early stopping scheme was applied as soon as the learning rate decreases twice in succession. As shown in Table \ref{dataset}, the \textbf{vox\_clean\_aug} is used to pre-train the speaker representation module. Only the cross entropy loss on the speaker label is used in this stage. We use the trained speaker representation module as one of our baseline models to show the effect of speech enhancement as well.

The trained speaker representation module is fixed as the auxiliary network to pre-train our speaker enhancement module (Pre-training 2). The enhancement module is optimized using the perceptual loss and cross entropy loss jointly. As summarized in Table \ref{dataset}, three pairs of clean and noisy utterance are used in this stage. For \textbf{vox\_noisy}, their clean data is the corresponding sample in \textbf{vox\_clean}. For \textbf{vox\_noisy\_aug}, two types of augmentation are included just like \textbf{vox\_clean\_aug}. When utterances in \textbf{vox\_noisy} are augmented with MUSAN, we expect our enhancement module to remove the effect of augmentation as well. So the clean data for these utterances are their corresponding utterances in \textbf{vox\_clean}. However, for the utterances in \textbf{vox\_noisy\_aug} which are convolved with RIRs, we don't expect our enhancement work for dereverberation. So their corresponding clean utterances should come from \textbf{vox\_clean\_aug} in which samples augmented with RIRs. 
The batch size of this stage is 128, which includes 64 noisy utterances and their clean version. We use the Adadelta optimizer and initial learning rate of 0.3 in this stage. The learning rate decrease and early stopping scheme are identical with Pre-training 1. 

After pre-training, the speech enhancement module and speaker representation module are finetuned together to reduce potential mismatch. Except the initial rate is 0.0001 in this stage, both training setting and data set are same as what we used in Pre-training 2.

\subsection{Evaluation}
To evaluate the performance of our system to extract robust speaker representation in both clean and noisy environments, we conduct the speaker verification task using both SITW and NSITW.
In the inference stage, \textbf{vox\_noisy\_aug} passing through the enhancement module is used to train the PLDA backend. The Equal Error Rate (EER) and minimum Detection Cost Function (minDCF) with target prior $p = 0.05$ are used to evaluate our system.

\subsection{Baseline}
As we mentioned in Section \ref{training}, one of our baseline models is our proposed model without speech enhancement module (PL-EESR w/o enh), which is used to compare the effect of speech enhancement on our system. 
Then we use the standard x-vector as our second baseline model (x-vector). We trained the model using the Kaldi SITW/v2 recipe. We set the x-vector as the state-of-the-art model for the speaker verification task. We trained PL-EESR w/o enh using the setting of Pre-training 1 which is introduced in Section \ref{training}. This is the difference between it and x-vector.

\section{Results}
  \begin{table}[!t]
    \caption{Comparison of our PL-EESR with two baseline models on SITW and NSITW.}
  \label{result1}
  \centering
  \setlength\tabcolsep{3pt}
  
  \begin{tabular}{ p{1cm}ccclcc}

    \toprule
    \multicolumn{1}{c}{Test set}& \multicolumn{1}{c}{System}& \multicolumn{2}{c}{Dev} &  & \multicolumn{2}{c}{Eval} \\ \cline{3-4} \cline{6-7} 
   & & EER & DCF & &EER&DCF\\
   \hline
   \multirow{3}*{SITW}& x-vector&  2.965&0.1946 & &3.417 & 0.2241\\
   &PL-EESR w/o enh &3.656&0.2215&&3.442 &  0.2210\\

   & \textbf{PL-EESR}  &\textbf{2.851} & \textbf{0.1805}& &\textbf{2.460}& \textbf{0.1690}\\
   \hline
   \multirow{3}*{NSITW}
   & x-vector& 6.816&0.3685 & &7.190& 0.4131\\
   &PL-EESR w/o enh &7.470&0.3835&&  8.066& 0.4550\\

   & \textbf{PL-EESR}  &\textbf{4.698} & \textbf{0.3007}& & \textbf{5.272} &\textbf{0.3334} \\
    \bottomrule
  \end{tabular}

\end{table}

  \begin{table*}[h]
  \caption{Comparison of our end-to-end training scheme with different training settings. ``Joint" and ``Separate" denote whether the enhancement module and speaker representation module are fine-tuned together. ``Noisy" and ``Clean" denote whether the network is optimized on the gradient of noisy utterance or clean utterance.\\}
  \label{result2}
  \centering

  \begin{tabular}{ ccccccclcc}

    \toprule
    \multicolumn{1}{c}{Test set}& \multicolumn{1}{c}{System}& \multicolumn{1}{c}{Training}& \multicolumn{1}{c}{CE Loss Objective} &\multicolumn{1}{c}{Perceptual Loss Objective} &\multicolumn{2}{c}{Dev} &  & \multicolumn{2}{c}{Eval} \\ \cline{6-7} \cline{9-10} 
   &&& &&EER & DCF & &EER&DCF\\
   \hline
   \multirow{4}*{SITW}&(a)&Separate  &Noisy &- &\textbf{2.734}&0.1906  && 2.788 & 0.1833 \\
   &(b) &Separate&-&Noisy&3.812 & 0.2274& &3.827&0.2400 \\
   &(c) &Separate&Noisy; Clean&Noisy; Clean&2.811 &0.1919 & &2.570&0.1804  \\
   &(d)  &Joint&Noisy; Clean&Noisy; Clean& 2.851&\textbf{0.1805}& &\textbf{2.460}& \textbf{0.1690}\\
   \hline
   \multirow{4}*{NSITW}&(a)&Separate  &Noisy &- &5.930 &  0.3351&&5.931& 0.3706\\
   &(b) &Separate&-&Noisy&7.624 &0.3951& &8.124&0.4611\\
   &(c) &Separate&Noisy; Clean&Noisy; Clean&5.160 &0.3161 & &5.632& 0.3539\\
   &(d)  &Joint&Noisy; Clean&Noisy; Clean&  \textbf{4.698}&\textbf{0.3007} & &\textbf{5.272}& \textbf{0.3334}\\
    \bottomrule
  \end{tabular}

\end{table*}
\subsection{Baseline results}
Table \ref{result1} shows the comparison of the baselines and our PL-EESR on the development set and evaluation set of SITW (top) and NSITW (bottom) respectively. The text in bold is the best performance for each metric in different conditions. We can find in clean condition, compared without enhancement processing, our method achieves 28.5\% and 23.5\% relative improvements in terms of EER and minDCF respectively in evaluation set, 22.0\% and 18.5\% in the development set. In a noisy environment, the relative improvements in terms of EER and minDCF are 34.6\% and 26.7\% for the evaluation set, 37.1\% and 21.6\% for the development set. Compared to the standard x-vector, which can be seen as the state of the art model for speaker representation, the relative improvements are 28.0\% and 24.6\% in terms of EER and minDCF for evaluation set, 3.8\% and 7.2\% for the development set for SITW. For NSITW, the improvements in evaluation set and development set are 26.7\%, 19.3\% and 31.1\%, 18.4\% respectively.

We can conclude from this comparison that our proposed method is beneficial to speaker representation extraction from both clean and noisy utterances.

\subsection{Overall comparisons}
Table \ref{result2} shows the performance of our end-to-end model with different training settings. Firstly, we trained the speech enhancement module and speaker verification module separately to verify the effect of the perceptual loss, which means there is no finetune stage in System (a)-(c) in Table \ref{result2} and all of these three systems are optimized in the stage of Pre-training 2. In the first experiment, we use only the feedback of the speaker classification task, which is the cross entropy loss on predicted speaker ID with noisy utterance as input. The system flow chart is shown in \ref{exp}(a), and this training scheme is based on the basic idea in \cite{shon2019voiceid}. The results are summarized as System (a) in Table \ref{result2}. 

System (b) is the original perceptual loss used in \cite{kataria2020feature}. In this experiment, 
the clean utterance is set as the target and the module is trained on the gradient of noisy utterance as shown in \ref{exp} (b). This performance is summarized in System (b) in Table \ref{result2}. 

Then we modified the perceptual loss function. Since the cross entropy loss on the speaker label can be used to enlarge the inter-class distance, the perceptual loss in our system focus on decrease the intra-class distance. Specifically, after computing the perceptual loss, the module is optimized on both the gradient of noisy utterance and clean utterance. To make sure this idea works for utterance in clean environments as well, the cross entropy is computed for both noisy utterance and clean utterance. The flow chart of this experiment is shown in Fig. \ref{exp} (c) and results are summarized in System (c) of Table \ref{result2}.

System (d) is used to verify the facility of training jointly to decrease the mismatch between these two modules. The only difference between System (c) and System (d) in Table \ref{result2} is that System(d) is trained in finetune stage.

Comparing the results of System (a)-(c) in Table \ref{result2} with baseline performance in Table \ref{result1}, we can find using System (b) harmful for both clean and noisy environments. Perhaps this degradation is because the \textbf{vox\_clean} still involved some background noise although we set it as the target in training.
However, both System (a) and System (c) outperform the baseline. This shows the necessity of end-to-end speech enhancement to reduce distortion and information loss. 
 Finally, the System (d) gains the most improvements in all of these four systems in most conditions except the EER of devaluation set of SITW.

\section{Conclusion}
Motivated by the unsatisfactory performance of speech enhancement applied on speaker embedding extraction task, we proposed the end-to-end training scheme for robust speaker representation. The model is trained on loss functions that aim at mapping the noisy and clean utterances to the identical representation as well as classifying speaker labels. The experiment results show that our method outperforms baseline models in both clean and noisy environments.
\bibliographystyle{IEEEbib}
\bibliography{strings,refs}

\end{document}